\pdfoutput=1
\documentclass[prx,twocolumn,superscriptaddress]{revtex4-1} 

\usepackage{graphicx}
\usepackage{amsmath}
\usepackage{amssymb}
\usepackage[colorlinks=true, linkcolor=blue, urlcolor=blue,
            citecolor=blue]{hyperref}
\usepackage{hyperref}
\usepackage{bm}

\begin{document}
\title{Spintronics meets density matrix renormalization group: Quantum spin torque driven nonclassical magnetization reversal and dynamical buildup of long-range entanglement}

\author{Marko D.~Petrovi\'c}
\affiliation{Department of Physics and Astronomy, University of Delaware, Newark DE 19716, USA}

\author{Priyanka Mondal}
\affiliation{Department of Physics and Astronomy, University of Delaware, Newark DE 19716, USA}

\author{Adrian E.~Feiguin}
\affiliation{Department of Physics, Northeastern University, Boston, MA 02115, USA}

\author{Petr Plech\'a\v{c}}
\affiliation{Department of Mathematical Sciences, University of Delaware, Newark, DE 19716, USA}          

\author{Branislav K.~Nikoli\'c}
\email{bnikolic@udel.edu}
\affiliation{Department of Physics and Astronomy, University of Delaware, Newark DE 19716, USA}

\begin{abstract}
	We introduce time-dependent density matrix renormalization group (tDMRG) as a solution to {\em long standing} problem in spintronics---how to describe spin-transfer torque (STT) between flowing spins of conduction electrons and localized spins within a magnetic material by treating the dynamics of both spin species {\em fully}  quantum-mechanically. In contrast to conventional Slonczewski-Berger STT, where the localized spins are viewed as classical vectors obeying the Landau-Lifshitz-Gilbert equation and where their STT-driven dynamics is initiated {\em only} when the spin-polarization of flowing electrons and localized spins are {\em noncollinear}, quantum STT can occur when these vectors are {\em collinear but antiparallel}.  Using tDMRG, we simulate the time evolution of a many-body quantum state of electrons and localized spins, where the former are injected as a spin-polarized current pulse while the latter comprise a quantum Heisenberg ferromagnetic metallic (FM) \mbox{spin-$\frac{1}{2}$} XXZ chain initially in the ground state with spin-polarization  antiparallel to that of injected electrons. The quantum STT reverses the direction of localized spins, but without rotation from the initial orientation, when the number of injected electrons exceeds the number of localized spins. Such {\em  nonclassical reversal}, which is absent from LLG dynamics,  is strikingly inhomogeneous across the FM chain and it can be accompanied by reduction of the magnetization associated with localized spins, even to zero at specific locations. This is because quantum STT  generates a highly entangled nonequilibrium many-body state of all flowing and localized spins, despite starting from the initially unentangled ground state of a mundane FM. Furthermore, the mutual information between localized spins at the FM edges remains nonzero even at infinite separation as the signature of dynamical buildup of {\em long-range} entanglement. The growth-in-time of entanglement  entropy differentiates between the quantum and conventional (i.e., noncollinear) setups for STT, reaching much larger asymptotic value in the former case.
\end{abstract}

\maketitle

\section{Introduction}\label{sec:intro}

\begin{figure}[h!]
	\includegraphics[scale=1.0]{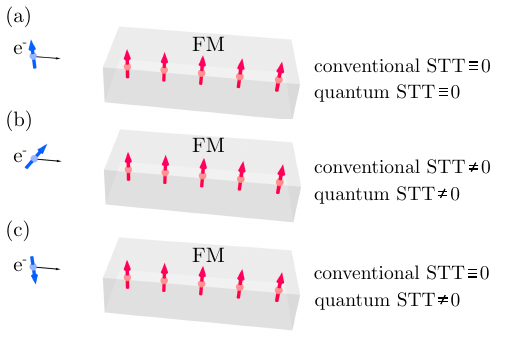}
	\caption{Illustration of three types of geometries of flowing conduction electron spins (blue arrow), assumed to be polarized by FM-polarizer layer (not shown explicitly), with respect to localized spins (red arrows) within FM-analyzer layer onto which electrons are impinging: (a) parallel; (b) noncollinear; and (c) antiparallel. The conventional STT~\cite{Slonczewski1996,Berger1996,Ralph2008} is nonzero only in (b), while quantum STT is nonzero in both (b) and (c). Blue and red arrows represent expectation values of the corresponding quantum-mechanical spin operators. For conventional STT, red arrows are modeled as classical vectors of fixed length~\cite{Ralph2008,Petrovic2018,Berkov2008}, whose time evolution due to conventional STT is animated by three TDNEGF+LLG-computed movies in the SM~\cite{sm} [showing no dynamics for panels (a) and (c)].}
	\label{fig:fig0}
\end{figure}

The conventional spin-transfer torque (STT) has been at the forefront of basic~\cite{Ralph2008} and applied~\cite{Locatelli2014} research in spintronics since the seminal theoretical predictions of Slonczewski~\cite{Slonczewski1996} and Berger~\cite{Berger1996}. Its {\em key requirement} is that the spin-polarization of flowing conduction electrons injected into a  ferromagnetic metal (FM) must be {\em noncollinear} to FM magnetization, as illustrated in Fig.~\ref{fig:fig0}(b). Thus,  it came as a great surprise when current-driven magnetization dynamics was recently observed at ultralow $T \sim 1$ K temperatures~\cite{Zholud2017,Zhang2017} in spin valves  FM-polarizer/normal-metal/FM-analyzer with {\em collinear} magnetizations.  Although thermal fluctuations of magnetization can create the required noncollinearity in spin valves (or magnetic tunnel junctions) at room temperature~\cite{Zhang2017}, they are frozen at ultralow temperatures of the experiment in Ref.~\cite{Zholud2017}. Thus, the effect observed in Ref.~\cite{Zholud2017} was dubbed ``quantum STT''~\cite{Zhang2017} and believed to be dissociated from conventional STT. In fact, few earlier experiments~\cite{Balashov2008,Kim2016,Kim2019} have reported  current-driven excitation of high energy magnons (with $\sim 1$ THz frequencies, which is orders of magnitude higher than typical $\sim  1$ GHz  magnetization dynamics driven by conventional STT), suggesting that collinear [but antiparallel, as illustrated in Fig.~\ref{fig:fig0}(c)] spin-polarization of flowing conduction electrons and localized magnetic moments drives the dynamics of the latter which is, therefore, also apparently dissociated from conventional STT.

The ``standard model''~\cite{Ralph2008} of conventional STT involves localized magnetic moments $\mathbf{M}_i$, viewed as {\em classical vectors of fixed length}, which interact with a nonequilibrium electronic spin density $\mathbf{s}_i$, computed by some steady-state~\cite{Wang2008b,Ellis2017,Belashchenko2019,Dolui2019} or time-dependent~\cite{Petrovic2018,Bajpai2019,Suresh2021} single-particle quantum transport formalism. The nonequilibrium electronic spin density is then fed 
into the Landau-Lifshitz-Gilbert (LLG) equation~\cite{Berkov2008} for $\mathbf{M}_i(t)$ in order to include Slonczewski-Berger STT $\propto \mathbf{s}_i \times \mathbf{M}_i$. Thus, in the context of collinear spin valve setup of Ref.~\cite{Zholud2017}, where the conventional Slonczewski-Berger STT $\propto \mathbf{s}_i \times \mathbf{M}_i \equiv 0$, the ``standard model'' predicts 
{\em no effect}. This is also illustrated by static $\mathbf{M}_i(t)$, despite injected current pulse, in the movies in the Supplemental Material (SM)~\cite{sm} animating Figs.~\ref{fig:fig0}(a) and ~\ref{fig:fig0}(c), which are obtained from time-dependent nonequilibrium Green function combined with LLG (TDNEGF+LLG)  simulations~\cite{Petrovic2018,Bajpai2019,Suresh2021} as an example of {\em quantum}-for-electrons--{\em classical}-for-localized-spins approach falling into the category of the ``standard model.'' In contrast, $\mathbf{M}_i(t)$ exhibit nontrivial dynamics in the TDNEGF+LLG-computed movie corresponding to noncollinear setup of Fig.~\ref{fig:fig0}(b), as expected from conventional STT $\propto \mathbf{s}_i \times \mathbf{M}_i  \neq 0$ being nonzero in this setup.

Let us recall that, in general, LLG description~\cite{Berkov2008} of the dynamics of localized spins is justified~\cite{Wieser2015,Wieser2016} only in the limit of large localized spins $S \rightarrow \infty$ and $\hbar \rightarrow 0$ (while $S \times \hbar  \rightarrow 1$), as well as in the  {\em absence of entanglement} in many-body quantum state of localized spins. Entanglement describes genuinely quantum and nonlocal correlations between different parts of a physical system.  While LLG description often captures experiments on realistic materials where $S$ is finite, it inevitably becomes inapplicable~\cite{Wieser2015,Wieser2016} in the presence of such many-body entanglement~\cite{Laflorencie2016,Chiara2018} because the length $|\mathbf{M}_i(t)|$ is then changing in time with smaller values signifying higher entanglement. For example, even if we start with a separable (unentangled) state of $N_\mathrm{FM}$ localized spins as the ground state of FM-analyzer at $t=0$, $|\Phi(t=0)\rangle_\mathrm{lspins} = |\!\! \downarrow_1 \downarrow_2 \cdots \downarrow_{N_\mathrm{FM}} \rangle$, spin-polarized current injection in the collinear setup of Fig.~\ref{fig:fig0}(c) or noncollinear setup of Fig.~\ref{fig:fig0}(b) eventually generates superpositions [Eq.~\eqref{eq:explicitket}] of such separable states so that quantum state of localized spins becomes both mixed~\cite{Elben2020a}  (due being subsystem of a larger total system which includes flowing electrons) and entangled with its measures of entanglement monotonically increasing in time [Fig.~\ref{fig:fig6}]. In general, nonequilibrium quantum systems left unobserved (i.e., without their unitary evolution being punctuated by nonunitary projective measurements) tend to evolve toward states of higher entanglement~\cite{Skinner2019}, as observed experimentally~\cite{Brydges2019} at sufficiently low temperature ensuring that decoherence due to external environment is suppressed. In addition, $|\Phi(t=0)\rangle_\mathrm{lspins}$ could be entangled from the outset as in the case  of strongly electron-correlated and/or exotic solid-state materials such as quantum antiferromagnets~\cite{Petrovic2021,Mitrofanov2021}, Mott insulators~\cite{Petrovic2021} and quantum spin liquids---in all three cases, many-body entanglement~\cite{Laflorencie2016,Chiara2018} in the ground state in equilibrium leads to \mbox{$\mathbf{M}_i(t=0) \equiv 0$} so that one again encounters a situation where the conventional Slonczewski-Berger STT \mbox{$\propto \mathbf{s}_i \times \mathbf{M}_i \equiv 0$} cannot be initiated. Thus, either due to entanglement already present in \mbox{$|\Phi(t=0) \rangle_\mathrm{lspins}$} or due to dynamical buildup of entanglement in time-dependent quantum state, classical LLG equation for localized spins becomes inapplicable. Instead, time evolution of localized spins {\em must} be treated quantum-mechanically with their individual expectation values $\mathbf{S}_i(t)$ [or \mbox{$\mathbf{M}_i(t) \propto \mathbf{S}_i(t)$}] {\em calculated only at the end---we term any such situation where the current-driven dynamics of localized spins must be described fully quantum-mechanically as quantum STT}.

Surprisingly, despite a long history of STT, an established fully quantum-mechanical framework for coupled dynamics of localized spins and flowing electron spins, as well as transfer of spin angular momentum between them, is still lacking~\cite{Zholud2017,Zhang2017,Tay2013}. Since both electrons and localized spins have to be evolved quantum-mechanically by such framework, it invariably has to be constructed using the tools of nonequilibrium quantum many-body theory. A handful of recent theoretical studies~\cite{Qaiumzadeh2018,Bender2018,Mondal2019,Mitrofanov2021,Mitrofanov2020} have offered insights into possible microscopic mechanisms of quantum STT. However, they rely on either: ({\em i})  a mapping of original operators of localized spins to bosonic operators and additional approximations~\cite{Mahfouzi2014} that {\em do not} allow us to track the time evolution of localized spins once they deviate too far from the initial orientation set by the anisotropy axis~\cite{Qaiumzadeh2018,Bender2018}; or ({\em ii}) they consider {\em only one} injected spin-polarized electron~\cite{Mondal2019,Mitrofanov2021,Mitrofanov2020}, which is insufficient to reverse many localized spins because of demand posed by spin angular momentum conservation. 

\begin{figure}
	\includegraphics[width=0.48\textwidth]{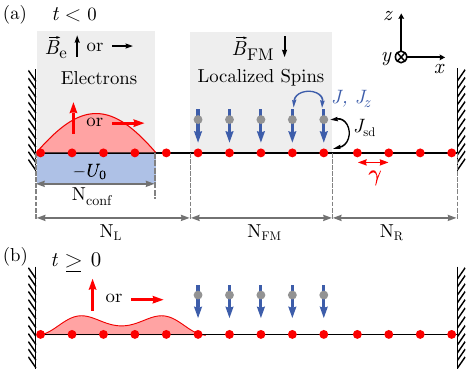}
	\caption{Schematic view of a two-terminal setup for tDMRG calculations where 1D tight-binding chain (blue dots) 
		of \mbox{$N = 75$} sites, with electron nearest-neighbor hopping $\gamma$ between all sites, hosts \mbox{$N_{\rm FM} = 5$} localized spins-$\frac{1}{2}$ (red  arrows) comprising a ferromagnetic quantum Heisenberg XXZ chain. First $N_{\rm L}=35$ sites within the left fermionic lead also include $N_{\rm conf} = 10$ sites where the confining potential $V$ is applied to $N_e \in \{ 1,5,8 \}$ electrons filling those sites. For $t < 0$, external magnetic fields $\mathbf{B}_e$ and $\mathbf{B}_{\rm FM}$ polarize electron spins along the $+z$-axis or $+x$-axis and localized spins along the $-z$-axis, respectively. For $t \ge 0$, both magnetic fields and the confining potential are switched off, so that electrons spread from left to right, as also animated by the tDMRG-computed movie in the SM~\cite{sm}.}
	\label{fig:fig1}
\end{figure}

In this study, we introduce the adaptive time-dependent density matrix renormalization group (tDMRG)~\cite{White2004,Daley2004,Feiguin2011,Paeckel2019} as 
a numerical framework capable of describing quantum and conventional STT on the same footing. Since this simulation method works directly with the original quantum-mechanical operators of the localized spins, it can capture reversal of localized spins due to STT which is highly sought in spintronic applications~\cite{Ralph2008,Locatelli2014,Dolui2019,Suresh2021}.  We demonstrate this by applying the tDMRG to a  one-dimensional (1D)  setup depicted in Fig.~\ref{fig:fig1} where quantum Heisenberg FM spin-$\frac{1}{2}$ XXZ chain is attached to the left (L) and right (R) fermionic leads~\cite{Lange2018,Lange2019}  modeled as 1D tight-binding chains of finite length. The {\em nonzero} electron hopping between the sites of the XXZ chain means that FM chain models {\em metallic} FM-analyzer layer that is receiving STT. From the viewpoint of the physics of strongly correlated electrons, this can also be interpreted as Kondo-Heisenberg chain~\cite{Tsvelik2017} sandwiched by fermionic leads, with ferromagnetic exchange interaction between localized 
spins, as well as between localized spins and injected flowing electrons. 

\begin{figure}
	\includegraphics[width=0.50\textwidth]{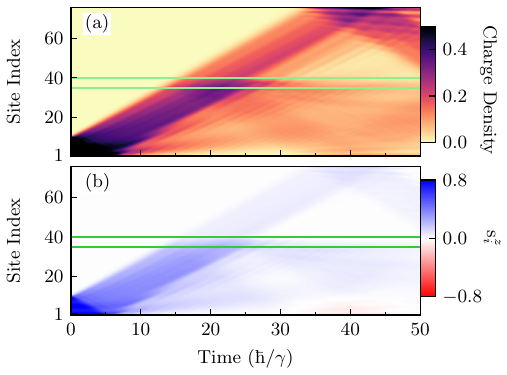}
	\caption{%
		Spatio-temporal profiles of electronic (a) charge density and (b) spin-$z$ density for spin-polarized current pulse composed of $N_e = 8$ electrons injected into the FM region in Fig.~\ref{fig:fig1}. The green horizontal lines in both panels mark the first and the last localized spin of the FM region. Electrons are initially ($t < 0$) spin-polarized along the $+z$-axis, while localized spins are polarized along the $-z$-axis. The strength of $sd$ exchange interaction between electron spin and localized spins is $J_\mathrm{sd}=0.5 \gamma$. Both panels are animated as the tDMRG-computed movie in the SM~\cite{sm} for $J_\mathrm{sd}=0.5 \gamma$ and $J_\mathrm{sd}=2.0 \gamma$.
		\label{fig:fig2}}
\end{figure}

The role of the FM-polarizer layer is simulated by filling the L lead with $N_e$ electrons (one per site), which are spin-polarized in a desired direction by applying an external magnetic field $\mathbf{B}_e$ [see Fig.~\ref{fig:fig1}(a) depicting the region where this field is applied] in that direction. They are also confined into a quantum well for times $t<0$, as illustrated in Fig.~\ref{fig:fig1}(a).  By removing the confining potential for times $t \ge 0$, electrons  spread into the region of the localized FM moments, as shown in Fig.~\ref{fig:fig2} and animated in the tDMRG-computed movie in the SM~\cite{sm}. This protocol mimics injection of a spin-polarized current pulse often employed in STT-operated spintronic devices~\cite{Ralph2008,Locatelli2014,Dolui2019,Suresh2021}. Prior to explaining our principal results in Figs.~\ref{fig:fig2}--\ref{fig:fig7} for the STT-driven quantum dynamics of the local magnetization across the FM chain, we first introduce useful concepts and necessary notation.  

\section{Model Hamiltonian}\label{sec:methods}
The setup illustrated in Fig.~\ref{fig:fig1} is a 1D chain of $N$ sites where electrons and localized spins are described by the Hamiltonian 
\begin{equation}\label{eq:hamiltoniansum}
  \hat{H} = \hat{H}_e + \hat{H}_{\rm lspins} + \hat{H}_{e\mathrm{-lspins}} + \hat{H}_{V,\mathbf{B}}(t<0).
\end{equation}
The tight-binding Hamiltonian for electrons
\begin{equation}\label{eq:hamiltonianelectrons}
  \hat{H}_e = -\gamma\sum_{i=1}^{N-1} 
            \left(
            \hat{c}^\dagger_{i\uparrow}
            \hat{c}_{i+1\uparrow}
            +
            \hat{c}^\dagger_{i\downarrow}
            \hat{c}_{i+1\downarrow}
            + \mathrm{h.c.}
            \right),
\end{equation}
operates on all $N=75$ sites, where $\hat{c}^\dagger_{i\sigma}$ ($\hat{c}_{i\sigma}$) creates (annihilates) an electron with spin $\sigma=\uparrow,\downarrow$ on site $i$. The nearest-neighbor (NN) hopping parameter \mbox{$\gamma=1$ eV} sets a unit of energy. Each site hosts one of the four possible electronic quantum states---empty $|0\rangle$, spin-up $|\!\! \uparrow\rangle = \hat{c}^\dagger_{i\uparrow}|0\rangle$, spin-down $|\!\! \downarrow \rangle = \hat{c}^\dagger_{i\downarrow}|0\rangle$, and doubly occupied  $|\!\! \uparrow\downarrow\rangle = \hat{c}^\dagger_{i\uparrow} \hat{c}^\dagger_{i\downarrow}|0\rangle$---from which one can construct $4^N$ many-body 
states that span the Fock space $\mathcal{F}_e$. The operators for the total number of electrons $\hat{N}_e=\sum_{i=1}^N \hat{n}_{i}$ and total electron spin along the $\alpha$-axis $\hat{s}^\alpha_e=\sum_{i=1}^N \hat{s}^z_{i}$ are given by sums of local (per-site) charge  and spin density operators, $\hat{n}_{i} = \sum_{\sigma=\{\uparrow,\downarrow\}}\hat{c}^\dagger_{i\sigma}\hat{c}_{i\sigma}$ and $\hat{s}^{\alpha}_{i}=\sum_{\sigma=\{\uparrow,\downarrow\}} \hat{c}^\dagger_{i\sigma} \hat{\sigma}^\alpha_{\sigma\sigma'} \hat{c}_{i\sigma'}$, respectively. Out of $N=N_{\rm L}+N_{\rm FM}+N_{\rm R}$ sites in Fig.~\ref{fig:fig1}, the first  $N_{\rm L}=35$ belong to the L fermionic lead and the last $N_{\rm R}=35$ belong to the R fermionic lead. The middle $N_{\rm FM}=5$ sites host localized spins whose mutual interaction is described by ferromagnetic XXZ spin-$\frac{1}{2}$ quantum Heisenberg Hamiltonian
\begin{equation}\label{eq:hamiltonianlspins}
   \hat{H}_{\rm spins} = -
       \sum_{i=1}^{N_{\rm FM}-1}
          \left[J_z
          \,\hat{S}^z_i \cdot \hat{S}^z_{i+1} 
         + J\left(\hat{S}^x_i\cdot\hat{S}^x_{i+1} 
                 + \hat{S}^y_i\cdot\hat{S}^y_{i+1}\right)
                 \right]. 
\end{equation}
Here $\hat{\mathbf{S}}_i$ is the spin-$\frac{1}{2}$ operator located on lattice site $i$; and the NN exchange interactions between localized spins are $J = 0.1\gamma$ and $J_z = 0.1005\gamma$, thereby including anisotropy along the $z$-axis. The $2^{N_\mathrm{FM}}$-dimensional Hilbert space of all localized spins is constructed as \mbox{$\mathcal{H}_{\rm lspins} = \mathcal{H}_1 \otimes \mathcal{H}_2 \otimes \cdots \otimes \mathcal{H}_{N_{\rm FM}}$}. Thus, the total Hamiltonian in Eq.~\eqref{eq:hamiltoniansum} acts on the space $\mathcal{F}_e \otimes \mathcal{H}_\mathrm{lspins}$, where the interaction between conduction electron spins and localized spins is described by
\begin{equation}\label{eq:hamiltoniane-lspins}
 \hat{H}_{e\mathrm{-lspins}} = -
    \sum_{i=N_\mathrm{L}+1}^{N_{\rm FM}}
    J_\mathrm{sd}
    \left(
    \hat{s}^x_{i}\!\cdot\!\hat{S}^x_{i} 
    +
    \hat{s}^y_{i}\!\cdot\!\hat{S}^y_{i}
    +
    \hat{s}^z_{i}\!\cdot\!\hat{S}^z_{i} 
    \right). 
\end{equation}
Here $J_\mathrm{sd}=0.5 \gamma$ (the tDMRG-computed movie in the SM~\cite{sm} shows additional case with $J_\mathrm{sd}=2.0 \gamma$) is interpreted as either $sd$~\cite{Ralph2008} or Kondo ferromagnetic exchange~\cite{Tsvelik2017} interaction in the fields of spintronics or strongly correlated electrons, respectively.

\begin{figure}
	\includegraphics[width=0.48\textwidth]{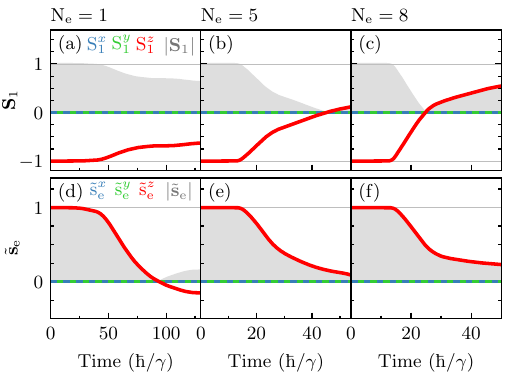}
	\caption{Time evolution of the expectation value of first localized spin $\mathbf{S}_{\rm 1} = (S_{\rm 1}^x, S_{\rm 1}^{y}, S_{\rm 1}^{z})$ and the purity $|\mathbf{S}_{1}|$ of its quantum state (gray background) for a different number of injected electrons which are initially spin-polarized along the $+z$-axis: (a) $N_{e} = 1$; 
		(b) $N_{e} = 5$; and (c) $N_{e} = 8$. Panels (d)--(f) plot average electron spin expectation value, $\tilde{\mathbf{s}}_e = \mathbf{s}_e/ N_e$, and purity, $|\tilde{\mathbf{s}}_e|$. The $sd$ exchange interaction between electron spin and localized spins is $J_\mathrm{sd}=0.5 \gamma$. Panels (c) and (f) are animated as the tDMRG-computed movie in the SM for $J_\mathrm{sd}=0.5 \gamma$ and $J_\mathrm{sd}=2.0 \gamma$.}
	\label{fig:fig3}
\end{figure}

For the purpose of preparing a many-electron spin-polarized current pulse, we employ the following term
\begin{eqnarray}\label{eq:hamiltonianconfine}
  \lefteqn{\hat{H}_\mathrm{V,\mathbf{B}}(t<0)  =  -V \sum_{i=1}^{N_\mathrm{conf}}
  \left(
  \hat{c}^\dagger_{i\uparrow}\hat{c}_{i\uparrow}
  +
  \hat{c}^\dagger_{i\downarrow}\hat{c}_{i\downarrow}
  \right)} \nonumber \\ 
  & & \mbox{}
  -\sum_{i=1}^{N_{\rm conf}} 
      g \mu_\mathrm{B} \hat{{\mathbf{s}}}_{i}\cdot\mathbf{B}_e 
  -\sum_{i=N_\mathrm{L}+1}^{N_{\rm FM}} 
      g \mu_\mathrm{B} \hat{\mathbf{S}}_{i}\cdot\mathbf{B}_{\rm FM},
\end{eqnarray}
 in Eq.~\eqref{eq:hamiltoniansum} which acts at times $t<0$ and is used only once to initialize the system. The first term in Eq.~\eqref{eq:hamiltonianconfine} is a confining on-site potential of magnitude $V=10 \gamma$ acting within the first $N_\mathrm{conf}=10$ sites of $N_\mathrm{L}=35$ sites of the L fermionic lead, as illustrated in Fig.~\ref{fig:fig1}(a). In addition, the second term in Eq.~\eqref{eq:hamiltonianconfine} polarizes, via an external magnetic field $|g \mu_\mathrm{B}  \mathbf{B}_e| = 100 \gamma$, the confined electrons along the $+z$-axis for the collinear setup of quantum STT analyzed in Figs.~\ref{fig:fig2}, ~\ref{fig:fig3}, ~\ref{fig:fig5}(a),(b), ~\ref{fig:fig6} and ~\ref{fig:fig7}, as well as in the tDMRG-computed movie in the SM~\cite{sm}; or spin-polarizes them along the $+x$-axis for the noncollinear setup of conventional STT~\cite{Ralph2008} analyzed in Figs.~\ref{fig:fig4},  ~\ref{fig:fig5}(c)--(e) and ~\ref{fig:fig6}. The third term in Eq.~\eqref{eq:hamiltonianconfine} is employed to polarize the localized spins along the $-z$-axis using an external magnetic field $|g \mu_\mathrm{B} \mathbf{B}_{\rm FM}|=100 \gamma$. The electron gyromagnetic ratio is denoted by $g$, and $\mu_\mathrm{B}$ is the Bohr magneton.

\section{tDMRG methodology adapted to quantum spin-transfer torque}\label{sec:tdmrg}

The exact time evolution of the system in Fig.~\ref{fig:fig1} can, in principle, be obtained by brute force application of the evolution operator
\begin{equation}\label{eq:evolutionop}
	|\Psi(t+\delta t) \rangle = e^{-i\hat{H} \delta t/\hbar} |\Psi(t)\rangle.
\end{equation}
Such an approach is, however, limited to small systems due to the exponential increase of the basis with system size. For example, for a system of $N=75$ sites hosting $N_\mathrm{FM}=5$ localized spin-$\frac{1}{2}$, onto which spin-polarized current pulse composed of $N_e=8$ electrons is impinging in Fig.~\ref{fig:fig1}(b), the vectors and matrices in Eq.~\eqref{eq:evolutionop} have size  $\binom{2N}{N_e}2^{N_\mathrm{FM}} \approx 1.68 \times 10^{14}$. 

To overcome this unfavorable scaling, we employ adaptive tDMRG~\cite{White2004,Daley2004,Feiguin2011,Paeckel2019} for which computational complexity is polynomial (instead of exponential) in system size. Let us first recall that the ground state DMRG~\cite{White1992,White1993,Schollwock2005} method can provide extremely accurate results for a many-body Hamiltonian [such as $\hat{H}$ in Eq.~\eqref{eq:hamiltoniansum}]. The premise is to obtain a wavefunction that approximates the actual ground state in a reduced Hilbert space. The proposed solution has the very peculiar form of a ``matrix-product state'' (MPS)~\cite{Oestlund1995}
\begin{eqnarray}\label{eq:mps}
	|\Psi\rangle & = & \sum_{\{s\}}A[s_1]_{\alpha_1}A[s_2]_{\alpha_1,\alpha_2}...A[s_{N-1}]_{\alpha_{N-1}\alpha_N}A[s_N]_{\alpha_N}|s_1...s_N\rangle 
\end{eqnarray}
where the coefficients of an MPS are generated by contracting matrices $A$ that are identified by a label corresponding to the state of the physical degree of freedom (the spin $s$, for instance). The row and column indices of the matrices correspond to the so-called ``bond indices'', with a ``bond dimension'' $\chi$, also referred to as the number of DMRG basis states. One has to find the coefficients of this wavefunction variationally, and the DMRG is one way to do it efficiently. The accuracy of the wavefunction increases with the bond dimension, and can be made asymptotically exact as this bond dimension approaches the total number of degrees of freedom. Most importantly, no {\it a priori} assumptions are made about the form of the coefficients, or the underlying physics. The power of the method is precisely that it is``smart'' enough to be able to find for us the best possible candidate wavefunction of that form. Moreover, it can find numerically exact results (within machine precision) even with small matrices (small bond dimension). Even though the accuracy is finite, it is  under control, so that we can obtain results that are essentially exact by just increasing the matrix size.

The generalization of DMRG to time-dependent problems requires to iteratively optimize the matrices, which is known as adaptive tDMRG algorithm, such that the balanced least-squares representation of the wavefunction is achieved for the whole time interval of propagation. We use the adaptive tDMRG formulation of Ref.~\cite{White2004} where the small-time-evolution 
operator is decomposed into
\begin{equation}\label{eq:decompose}
	e^{-i\hat{H} \delta t/\hbar}  \approx e^{-i\hat{H}_1 \delta t/2\hbar}  \cdots e^{-i\hat{H}_{N-1} \delta t/2\hbar} e^{-i\hat{H}_{N-1} \delta t/2\hbar} \cdots e^{-i\hat{H}_{1} \delta t/2\hbar }, 
\end{equation}
for an arbitrary many-body Hamiltonian $\hat{H} = \sum_{i=1}^{N-1} \hat{H}_i$ with nearest neighbor interactions between $N$ sites and $\hat{H}_i$ denoting its term on the bond $i$. Such approximation incurs an error of the order $O(\delta t^3)$.  The small time step is chosen as $\delta t =0.1 \hbar/\gamma$. We start the propagation with $\chi = 100$ states and limit the truncation error to $10^{-7}$, while the maximal number of states allowed during the evolution is set to $\chi_\mathrm{max} = 400$. 


For $t \ge 0$, $\hat{H}_\mathrm{V,\mathbf{B}} \equiv 0$ so that spin-polarized conduction electrons spread out from the region of $N_\mathrm{conf}$ sites and are injected into the FM chain. This process is illustrated schematically in Fig.~\ref{fig:fig1}(b), while the local charge and spin-$z$ densities are computed numerically  in Fig.~\ref{fig:fig2} and animated in the tDMRG-computed movie in the SM~\cite{sm}. Since fermionic leads are not semi-infinite as in the usual single-particle quantum transport calculations~\cite{Wang2008b,Ellis2017,Belashchenko2019,Dolui2019,Petrovic2018,Bajpai2019,Suresh2021}, the many-body system composed of conduction electrons and localized spins can be evolved only for a limited time~\cite{Lange2018,Lange2019} before electrons are backscattered by the right boundary which breaks L$\rightarrow$R current flow. For example, in Fig.~\ref{fig:fig2} such backscattering occurs at $t \simeq 40 \hbar/\gamma$ for $N_e = 8$ injected electrons. Nevertheless, the quantum dynamics of flowing electron spins and localized spins captured by tDMRG before the boundary reflection is fully equivalent to that in an open quantum system.

\section{Quantum spin-transfer torque in collinear geometry}
In the collinear setup~\cite{Zholud2017,Zhang2017} of quantum STT, the spin-polarization of the injected conduction electrons is collinear but antiparallel to that of the localized spins at $t=0$. In the Fock space sector with zero electrons \mbox{$N_e=0$}, the many-body quantum state $|\Psi(t)\rangle$ for $t \ge 0$ within $\mathcal{F}_e \otimes \mathcal{H}_\mathrm{lspins}$ space is trivially  \mbox{$|\Psi(t) \rangle=|\mathrm{vac} \rangle_e \otimes |\mathrm{\Phi} \rangle_\mathrm{lspins}$} where the first factor of such separable quantum state is the electron vacuum state $|\mathrm{vac} \rangle_e \in \mathcal{F}_e$ and the second factor \mbox{$|\Phi \rangle_\mathrm{lspins} = |\!\! \downarrow_1 \ldots \downarrow_{N_\mathrm{FM}} \rangle \in \mathcal{H}_\mathrm{lspins}$} is the ground state of the FM chain. The Fock space sector $N_e=1$ has been studied for an infinite ($N_\mathrm{FM} \rightarrow \infty$) metallic FM chains long before~\cite{Shastry1981} theoretical predictions for STT, but with the focus on magnetic polarons as the bound state of the injected electron and low-energy excitations (spinons or magnons) of all localized spins. In such a case, and for a FM chain~\cite{Mondal2019,Mitrofanov2020} of finite length, we find $|\Psi(t \ge 0)\rangle = c_0(t)|\mathrm{orb} \rangle \otimes |\!\! \uparrow_e \rangle  \otimes |\mathrm{\Phi} \rangle  + c_1(t)|\mathrm{orb} \rangle \otimes |\!\! \downarrow_e \rangle \otimes  |\!\! \uparrow_1 \ldots \downarrow_{N_\mathrm{FM}} \rangle + \cdots + c_{N_\mathrm{FM}}(t) |\mathrm{orb} \rangle \otimes |\!\! \downarrow_e \rangle \otimes |\!\! \downarrow_1 \ldots \uparrow_{N_\mathrm{FM}} \rangle$. This superposition is constructed by including {\em all possible states} allowed by the conservation of the $z$-component of total spin,
\begin{equation}\label{eq:spinconservation}
[\hat{H},\hat{s}_e^z + \hat{S}_\mathrm{lspins}^z]=0,
\end{equation}
where $\hat{S}_\mathrm{lspins}^z = \hat{S}_1^z + \cdots \hat{S}_{N_\mathrm{FM}}^z$. Here $|\mathrm{orb} \rangle$ is orbital state of a single injected electron, and the coefficients $c_0(t), \ldots, c_{N_\mathrm{FM}}(t)$ studied in Ref.~\cite{Mondal2019} can be much more complicated than those for magnons (or spinons) in an infinite FM chain~\cite{Shastry1981}.

\begin{figure}
	\includegraphics[width=0.49\textwidth]{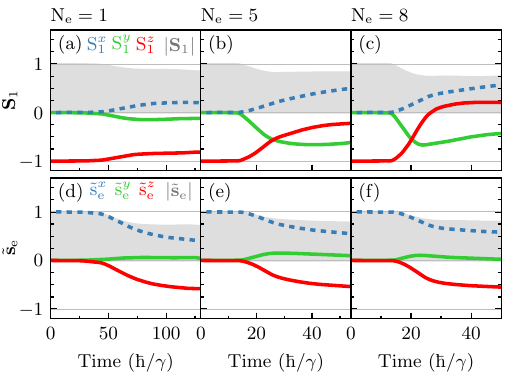}
	\caption{Panels (a)--(f) are counterparts of Fig.~\ref{fig:fig3}(a)--(f), but for injected electrons which are spin-polarized at $t<0$ along the $+x$-axis. This creates a noncollinear geometry of flowing and localized spins, as required for conventional STT~\cite{Ralph2008}.}
	\label{fig:fig4}
\end{figure}

The quantum state $|\Psi(t) \rangle$ also defines the pure state density matrix \mbox{$|\Psi(t) \rangle \langle \Psi(t)|$}. Since such state for $N_e \ge 1$ is a sum of separable states and, therefore, entangled, the quantum state of subsystems  must be described by the reduced density matrix~\cite{Laflorencie2016,Chiara2018,Elben2020a}. This is exemplified by  
\begin{equation}\label{eq:reducedrho}
\hat{\rho}_{1} = \mathrm{Tr}_\mathrm{other}\, |\Psi(t) \rangle \langle \Psi(t)| =  \frac{1}{2}\left[\hat{I} +
\mathbf{S}_{1} \cdot{\hat{\bm \sigma}} \right],
\end{equation}
which is the density matrix of the first localized spin (at site $N_\mathrm{L}+1$ in Fig.~\ref{fig:fig1}), obtained by partial trace over all states within $\mathcal{F}_e \otimes \mathcal{H}_\mathrm{lspins}$ that are not in $\mathcal{H}_1$. Here  $\hat{I}$ is the unit $2\times 2$ matrix and $\hat{\bm \sigma} = \left(\hat{\sigma}_x, \hat{\sigma}_y, \hat{\sigma}_z\right)$  is the vector of the Pauli matrices. The magnitude $|\mathbf{S}_{1}|$ of the expectation value of localized spin-$\frac{1}{2}$, 
\begin{equation}\label{eq:s1}
\mathbf{S}_{1} =\mathrm{Tr}\,[\hat{\rho}_{1}\hat{\bm \sigma}],
\end{equation}
also serves as {\em purity} specifying whether its quantum state  is fully ($|\mathbf{S}_{1}|=1$) or partially ($0<|\mathbf{S}_{1}|<1$) coherent.  We use label $O \equiv \langle \hat{O} \rangle$ for the expectation value of an operator $\hat{O}$ in a pure many-body state of the total system electrons plus localized-spins or in a mixed quantum state of a relevant (depending on observable $\hat{O}$) subsystem. Thus, {\em true} decoherence (i.e., decoherence that cannot be attributed to any classical noise~\cite{Kayser2015}) due to many-body entanglement~\cite{Laflorencie2016,Chiara2018} can lead to reduction of local and total magnetization, $\mathbf{M}_i=g\mu_{\rm B} \mathbf{S}_i$ and  $ \mathbf{M} = \sum_{i=N_\mathrm{L}+1}^{N_\mathrm{FM}} g\mu_{\rm B} \mathbf{S}_i$, respectively, because of  reduction of $\mathbf{S}_i$ expectation values. This is obviously forbidden in classical magnetization dynamics described by the LLG equation~\cite{Ellis2017,Petrovic2018,Berkov2008}.

The time evolution of $\mathbf{S}_{1}(t)$ is shown in Fig.~\ref{fig:fig3}(a)--(c) for  $N_e=1,5,8$ injected electrons, respectively; as well as in the tDMRG-computed movie in the SM~\cite{sm} for all $\mathbf{S}_{i}(t)$ using $N_e=8$. Due to spin angular momentum conservation, only $S_i^z(t) \neq 0$. The magnetization reversal sought in spintronic applications~\cite{Ralph2008,Locatelli2014}, where $S_i^z(t)$ evolves from $S_i^z =-1$ at $t=0$ to $S_i^z >0$ at some later time $t>0$, occurs only when  $N_e > N_{\rm FM}$. The reversal is {\em nonclassical} since $S_{i}^x(t) =  S_{i}^y(t) \equiv 0$, unlike classical magnetization reversal~\cite{Ralph2008,Dolui2019,Berkov2008} where $\mathbf{M}_i$ vectors must rotate away from the $-z$-axis to reach the $+z$-axis. The decoherence of localized spin states makes the reversal strikingly inhomogeneous (see the tDMRG-computed movie in the SM~\cite{sm}) because localized spins away from the L-lead/FM-chain interface have smaller $|\mathbf{S}_i|$ or $S_i^z$ can remain negative. The decoherence can be partially suppressed and all localized spins reversed by increasing $J_\mathrm{sd}$, despite larger $J_\mathrm{sd}$ concurrently enhancing reflection  of the current pulse at the L-lead/FM-chain interface (see the tDMRG-computed movie in the SM~\cite{sm}). The spin expectation value per electron, $\tilde{\mathbf{s}}_e =  \mathbf{s}_e /N_e$, plotted in Fig.~\ref{fig:fig3}(d)--(f) shows that, due to many-body entanglement, electron spin states also decohere with purity  $|\tilde{\mathbf{s}}_e|<1$.

\section{Quantum and conventional spin-transfer torque in noncollinear geometry}
As a comparison, we examine in Fig.~\ref{fig:fig4} conventional STT in a noncollinear geometry where injected electrons are spin-polarized along the $+x$-axis while localized spins are polarized along the $-z$-axis. Although this has been  considered~\cite{Zholud2017,Zhang2017} as a completely different situation from quantum STT in a collinear geometry, the state \mbox{$|\!\! \rightarrow^x_e \rangle$} in quantum language corresponds to the injection of a superposition of spin-up and spin-down states, \mbox{$|\!\! \rightarrow^x_e \rangle = (|\!\! \uparrow_e \rangle + |\!\! \downarrow_e \rangle)/\sqrt{2}$}. In this case, we find in Fig.~\ref{fig:fig4}(a)--(c) bow localized spins always rotate, $S_i^x \neq 0$  and $S_i^y \neq 0$, away from the easy $z$-axis for $t \ge 0$ akin to classical localized spins~\cite{Ralph2008,Dolui2019,Berkov2008}. However, $|\mathbf{S}_1|<1$ in Fig.~\ref{fig:fig4}(a)--(c) signifies the same decoherence due to many-body entanglement found for quantum STT in Fig.~\ref{fig:fig3}.

\begin{figure}
	\begin{center}
		\includegraphics[width=0.49\textwidth]{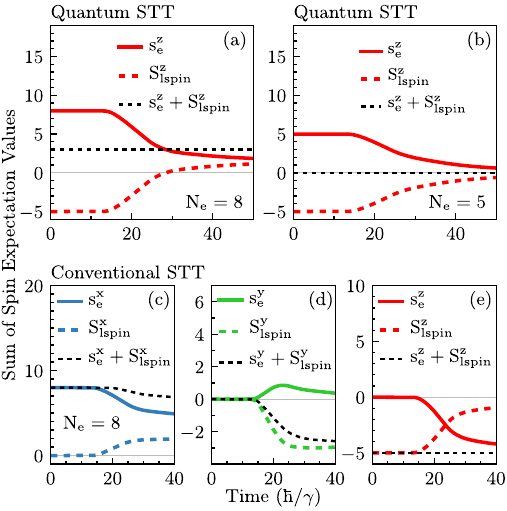}
	\end{center}
	\caption{Time evolution of the sum of spin expectation values of all $N_\mathrm{FM}=5$ localized spins $S_\mathrm{lspins}^z(t)$ and all injected (a) $N_e=8$ or (b) $N_e=5$  electrons $s_e^z(t)$ in collinear and antiparallel setup of quantum STT analyzed in Fig.~\ref{fig:fig3}(b),(c),(e),(f). The same time evolution, but for noncollinear setup of conventional STT using $N_e=8$ as analyzed in Fig.~\ref{fig:fig4}(c),(f), is shown in panels (c)--(e). The $z$-component of total spin is manifestly conserved (dashed blacked line) in panels (a), (b) and (e).}
	\label{fig:fig5}
\end{figure}

\section{What is ``transferred'' in spin-transfer torque?}
The conventional STT is commonly computed using some type of single-particle steady-state quantum transport formalism~\cite{Wang2008b,Ellis2017,Dolui2019} to obtain the nonequilibrium electron spin density $\mathbf{s}_i$ injected into the FM-analyzer. Due to noncollinearity between $\mathbf{s}_i$ and the classical magnetization $\mathbf{M}$ of the FM-analyzer, contributions to $\mathbf{s}_i$ from propagating states oscillate as a function of position without  decaying. Nevertheless, the transverse (with respect to $\mathbf{M}$) component of $\mathbf{s}_i$ is brought to zero within $\sim 1$ nm away from the normal-metal/FM-analyzer interface by averaging over propagating states with different incoming momenta $\mathbf{\hbar}\mathbf{k}$ because the  frequency of spatial oscillations rapidly changes with $\mathbf{k}$~\cite{Wang2008b}. The angular dependence of STT $\propto \sum_i \mathbf{s}_i \times \mathbf{M}$ can be fed~\cite{Ellis2017,Dolui2019} into the LLG calculations which often consider only the macrospin~\cite{Ralph2008,Berkov2008,Brataas2006a} $\mathbf{M}=\sum_i \mathbf{M}_i$. Thus, in this picture  the microscopic mechanism of how spin angular momentum is transferred from electron subsystem to magnetization remains hidden. 

The tDMRG simulations unveil such mechanism in Fig.~\ref{fig:fig5}(a),(b) for quantum STT, as well as in Fig.~\ref{fig:fig5}(c)--(e) for conventional STT, where the total spin of all electrons $s_e^z(t)$ decays in time while the total spin of all localized spins $S_\mathrm{lspins}^z(t)$ increases as injected flowing spins try to align localized spins in the same direction. Figure~\ref{fig:fig5}(a),(b),(e) also validates our calculations by confirming that $s_e(t) + S_\mathrm{lspins}^z(t)$ remains constant, as expected from the conservation law in Eq.~\eqref{eq:spinconservation}. Due to the complex superposition of many-body states of electrons plus localized-spins, the quantum dynamics of localized spins is always highly inhomogeneous and, therefore, quite different from  the macrospin approximation~\cite{Berkov2008} or simple spin wave excitations~\cite{Brataas2006a} assumed in the modeling of classical magnetization dynamics driven by conventional STT.

\begin{figure}
	\begin{center}
		\includegraphics[width=0.49\textwidth]{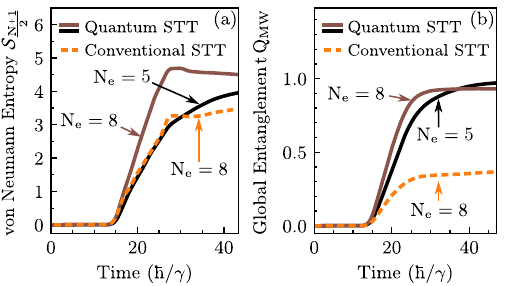}
	\end{center}
	\caption{(a) Time evolution of the von Neumann entropy [Eq.~\eqref{eq:entropy}] of half of the whole system in Fig.~\ref{fig:fig1}, which includes all electrons at time $t$ within the first 38 sites and three of localized spins within the FM region. (b) Time evolution of global entanglement measure [Eq.~\eqref{eq:mw}] for a subsystem composed of all conduction electron spins and all localized spins.}
	\label{fig:fig6}
\end{figure}

\section{Dynamical buildup of long-range entanglement}
\begin{figure}
	\begin{center}
		\includegraphics[width=0.49\textwidth]{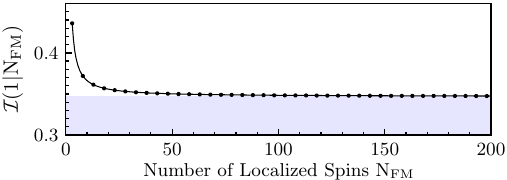}
	\end{center}
	\caption{Mutual information $\mathcal{I}(1|N_\mathrm{FM})$ [Eq.~\eqref{eq:mutual}]  between localized spins 1 and $N_\mathrm{FM}$ at the edges of FM region in Fig.~\ref{fig:fig1} as a function of its length $N_\mathrm{FM}$. The entangled nonequilibrium many-body state  characterized by  $\mathcal{I}(1|N_\mathrm{FM})$ is generated  by quantum STT exerted by $N_e=N_\mathrm{FM}$ injected electrons [with simplification that all possible terms in this state enter with equal weight in Eq.~\eqref{eq:explicitket}]. In the limit of infinite separation between the edges,  $N_\mathrm{FM} \rightarrow \infty \Rightarrow \mathcal{I}(1|N_\mathrm{FM}) \rightarrow \ln 2/2$.}		
	\label{fig:fig7}
\end{figure}

The nonequilibrium many-body states of electrons and localized spins generated by STT exhibit [Fig.~\ref{fig:fig6}] growth of entanglement entropy~\cite{Brydges2019,Bardarson2012}. Using entanglement measures~\cite{Laflorencie2016,Chiara2018} beyond entropy,   we also predict that they will 
exhibit long-range~\cite{Laflorencie2016,Chiara2018} entanglement [Fig.~\ref{fig:fig7}]. Massively and long-range entangled many-body quantum states have been sought among ground states of exotic phases of solid-state materials~\cite{Laflorencie2016,Chiara2018,Broholm2020} and synthetic quantum matter like Rydberg atoms and trapped ions~\cite{Elben2020}. In the latter case, entanglement growth has been measured experimentally~\cite{Brydges2019} in a system of $\sim 10$ trapped ion qubits. To quantify entanglement growth as a function of time, we compute the time evolution of the standard~\cite{Laflorencie2016,Chiara2018,Bardarson2012} von Neumann entanglement entropy for half of the system 
\begin{equation}\label{eq:entropy}
\mathcal{S}_\mathrm{\frac{N+1}{2}}(t)=-\mathrm{Tr}\, \hat{\rho}_\mathrm{\frac{N+1}{2}}(t) \ln \hat{\rho}_\mathrm{\frac{N+1}{2}}(t),
\end{equation}
where $\hat{\rho}_\mathrm{\frac{N+1}{2}}(t)$ is many-body density matrix of a subsystem composed of $3$ localized spins and of all electrons residing at time $t$ within first 38 sites of the system in Fig.~\ref{fig:fig1}. In addition, we also calculate the so-called  Meyer-Wallach (MW) measure~\cite{Chiara2018} of global entanglement~\cite{Chandran2007} which is defined for a multipartite quantum system composed of two-level subsystems as
\begin{equation}\label{eq:mw}
Q_\mathrm{MW} = 2\left(1 - \frac{1}{N_\mathrm{FM}+N_e} \left[\sum_{i=1}^{N_\mathrm{FM}} \mathrm{Tr}\, \hat{\rho}_i^2 + \sum_{j=1}^{N_e} \mathrm{Tr}\, \hat{\rho}_{e,j}^2 \right] \right). 
\end{equation}
It quantifies average entanglement of each subsystem with the remaining $N_\mathrm{FM}+N_e-1$ spins. The  nonequilibrium dynamics driven by quantum STT and local interactions in the Hamiltonian in Eq.~\eqref{eq:hamiltoniansum} conspire to increase both  $S_\mathrm{\frac{N+1}{2}}$ [Fig.~\ref{fig:fig6}(a)] and $Q_\mathrm{MW}(t)$ [Fig.~\ref{fig:fig6}(b)]. The latter stays slightly below its maximum possible value $Q_\mathrm{MW} = 1$ (obtained for $\mathrm{Tr}\, \hat{\rho}_i^2 = \mathrm{Tr}\, \hat{\rho}_{e,j}^2=0$) when $N_e>N_\mathrm{FM}$  because of the initial condition \mbox{$s_e^z + S_\mathrm{lspins}^z \neq 0$}.  Both $\mathcal{S}_\mathrm{\frac{N+1}{2}}(t)$ and $Q_\mathrm{MW}(t)$ reach smaller asymptotic value [Fig.~\ref{fig:fig6}] at longer times in the case of conventional STT in noncollinear geometry, so that they clearly differentiate between quantum and conventional STT.

For the purpose of demonstrating long-range entanglement in nonequilibrium quantum many-body state generated by quantum STT,  we additionally analyze the mutual information~\cite{Chiara2018}
\begin{equation}\label{eq:mutual}
\mathcal{I}(1|N_\mathrm{FM}) = \mathcal{S}_1 + \mathcal{S}_\mathrm{N_\mathrm{FM}}  - \mathcal{S}_{1,N_\mathrm{FM}},
\end{equation}
between localized spins at the edge of the FM region, i.e., at sites $1$ and $N_\mathrm{FM}$. Here $\mathcal{S}_1$ is the von Neumann entropy computed via Eq.~\eqref{eq:entropy} from the density matrix $\hat{\rho}_1$ [Eq.~\eqref{eq:reducedrho}] of localized spin $1$ at the left edge of FM;  $\mathcal{S}_{N_\mathrm{FM}}$ is the von Neumann entropy of localized spin at the right edge of the FM region; and  $\mathcal{S}_{1,N_\mathrm{FM}}$ is the von Neumann entropy of a subsystem composed of these two localized spins. The three entropies are evaluated for a many-body state generated after $N_e$ electrons are injected into FM with $N_\mathrm{FM}=N_e$ localized spins, so that at $t=0$ the state is separable, $|\Psi(t=0) \rangle =|\mathrm{orb}  \rangle \otimes |\!\! \uparrow_e \uparrow_e \cdots \uparrow_e \rangle  \otimes |\!\! \downarrow_1 \downarrow_2 \cdots \downarrow_{N_\mathrm{FM}} \rangle$. To show explicitly the type of state generated and also to be able to analyze its properties in the limit $N_\mathrm{FM} \rightarrow \infty$, we do not evolve initial state by tDMRG but instead write for $t>0$ 
\begin{eqnarray}\label{eq:explicitket}
|\Psi(t \ge 0) \rangle & = &  |\mathrm{orb} \rangle \otimes\frac{1}{\sqrt{C}} \Big(  |\!\! \uparrow_e \uparrow_e \cdots \uparrow_e \rangle  \otimes |\!\! \downarrow_1 \downarrow_2 \cdots \downarrow_{N_\mathrm{FM}} \rangle  \nonumber \\
&& + |\!\! \downarrow_e \uparrow_e \cdots \uparrow_e \rangle  \otimes |\!\! \uparrow_1 \downarrow_2 \cdots \downarrow_{N_\mathrm{FM}} \rangle + \ldots \nonumber \\ 
&& + |\!\! \downarrow_e \downarrow_e \cdots \uparrow_e \rangle  \otimes |\!\! \uparrow_1 \uparrow_2 \cdots \downarrow_{N_\mathrm{FM}} \rangle + \ldots \nonumber \\
&&  + |\!\! \downarrow_e \downarrow_e \cdots \downarrow_e \rangle  \otimes |\!\! \uparrow_1 \uparrow_2 \cdots \uparrow_{N_\mathrm{FM}} \rangle  \Big). 
\end{eqnarray}
The individual terms in this sum are {\em all possible} separable states {\em obeying} the spin conservation law in Eq.~\eqref{eq:spinconservation}, where we employ simplification where coefficients in front of each term are identical and time-independent. Our tDMRG simulation effectively generates proper nonuniform~\cite{Mondal2019}  time-dependent coefficients, and it can be conducted for  $N_\mathrm{FM}=N_e \sim 100$, but the state in Eq.~\eqref{eq:explicitket} can be written and analyzed for arbitrary large  $N_\mathrm{FM}$. There are $C=\binom{2N_\mathrm{FM}}{N_\mathrm{FM}} \sim 4^{N_\mathrm{FM}}/\sqrt{N_\mathrm{FM}}$ terms in the sum in  Eq.~\eqref{eq:explicitket}. Thus, the subspace of dimension $\sim 4^{N_\mathrm{FM}}/\sqrt{N_\mathrm{FM}}$ capturing time evolution of nonequilibrium states of the type in Eq.~\eqref{eq:explicitket} also furnishes an example where the majority of all possible $4^{N_\mathrm{FM}}$ states in the Hilbert space are unphysical in the sense of not being utilized in the course of time evolution~\cite{Poulin2011}. The von Neumann entropies of the edge localized spins, $\mathcal{S}_1=\mathcal{S}_{N_\mathrm{FM}+1}=1$, are obtained from $\hat{\rho}_1=\frac{1}{2}(|\!\!\uparrow_1\rangle \langle \uparrow_1\!\!| + |\!\!\downarrow_1\rangle \langle \downarrow_1 \!\!|)$ as incoherent mixture with zero off-diagonal elements, while 
\begin{eqnarray}\label{eq:entropyedgespins}
\mathcal{S}_\mathrm{1,N_\mathrm{FM}} & = & \frac{N_\mathrm{FM}}{2N_\mathrm{FM}-1} \ln \frac{2N_\mathrm{FM}-1}{N_\mathrm{FM}}  \nonumber  \\
&& + \frac{N_\mathrm{FM}-1}{2N_\mathrm{FM}-1} \ln \frac{4N_\mathrm{FM}-2}{N_\mathrm{FM}-1},
\end{eqnarray}
is obtained from Eq.~\eqref{eq:entropy} using $\hat{\rho}_{1,\mathrm{N}_\mathrm{FM}}=\mathrm{Tr}_\mathrm{other}|\Psi\rangle \langle \Psi|$ that contains also nonzero off-diagonal elements. The coherences encoded by the off-diagonal elements lead to nonzero mutual information in Fig.~\ref{fig:fig7} even at {\em infinite} separation between the edge spins
\begin{equation}\label{eq:limit}
\lim_{N_\mathrm{FM} \rightarrow \infty} \mathcal{I}(1|N_\mathrm{FM})=\frac{\ln 2}{2}, 
\end{equation}
as the signature of {\em long-range entanglement}. This demonstrates that pure nonequilibrium many-body states of the type displayed in Eq.~\eqref{eq:explicitket}  is {\em macroscopically} entangled and quantum correlated. Notice that this entanglement persists even as the electrons leave FM region and are no longer interacting with the localized spins, as demonstrated by the tDMRG-computed movie in the SM~\cite{sm}.

\section{Conclusions and Outlook}
In conclusion, we introduce tDMRG as fully quantum many-body framework for describing transfer of spin angular momentum between flowing electrons, comprising a current pulse, and localized spins. Unlike the ``standard model'' approaches to conventional Slonczewski-Berger STT~\cite{Ralph2008,Ellis2017,Petrovic2018,Bajpai2019,Suresh2021}, which are 
all based on single-particle quantum mechanics for electrons and classical LLG description of localized spins, tDMRG can describe spin transfer even when such approaches predict {\em completely absent} STT. This includes collinear but antiparallel localized and flowing spins in spin valves~\cite{Zholud2017} or in schemes exciting high energy magnons~\cite{Balashov2008,Kim2016,Kim2019}; as well as possible future experiments on spin-polarized current injection into strongly electron-correlated materials like quantum antiferromagnets~\cite{Petrovic2021,Mitrofanov2021} and Mott insulators~\cite{Petrovic2021} or exotic materials like quantum spin liquids where expectation value of localized spins is zero in equilibrium. In all of these situations, classical LLG equation description of localized spins is inapplicable due to many-body entanglement~\cite{Laflorencie2016,Chiara2018} in either equilibrium state or in nonequilibrium quantum many-body state (or both) of all flowing electrons and localized spins. The entanglement entropy of nonequilibrium quantum many-body state driven by quantum STT  grows in time [Fig.~\ref{fig:fig6}], while the state additionally  become long-ranged entangled [Fig.~\ref{fig:fig7}]. Thus, instead of LLG dynamics, such nonequilibrium quantum many-body state must be evolved and expectation values of localized spins computed only at the end---{\em we term any such situation quantum STT}. 

Looking to the future, modeling of quantum STT in two-dimensional~\cite{Dolui2019} and three-dimensional~\cite{Ellis2017} realistic spintronic device over  long times requires to develop many-body NEGF-based algorithms~\cite{Schlunzen2020} (as opposed to presently widely used single-particle NEGF-algorithms~\cite{Ellis2017,Belashchenko2019,Dolui2019,Petrovic2018,Bajpai2019,Suresh2021} applied to conventional STT) where a number of technical challenges~\cite{Mahfouzi2014} remains to be solved. For such necessarily perturbative efforts, our tDMRG approach to quantum STT offers rigorous nonperturbative benchmarking~\cite{Schlunzen2020} using one-dimensional examples like the one in Fig.~\ref{fig:fig1}.  

Although experimental measurement of many-body entanglement~\cite{Laflorencie2016,Chiara2018} has been achieved in cold gases of $\sim 10$ atoms using atomic-molecular-optical physics techniques~\cite{Brydges2019}, it remains an outstanding challenge~\cite{Laflorencie2016} for solid-state materials and devices. For the special case of quantum-STT-driven many-body entanglement in nonequilibrium spintronic devices studied here,  we propose that by injecting an electronic current pulse of sufficient magnitude, many-body entanglement of a macroscopically large number of flowing and localized spins can be detected: ({\em i}) by first  measuring spectrum of excitations via inelastic light scattering (Raman or Brillouin)~\cite{Arana2017} of FM-analyzer layer in equilibrium, where magnons peaks will be observed; ({\em ii}) immediately after the pulse has ceased, measure spectrum of excitations again where a broad continuum~\cite{Broholm2020} could be observed due to long-range entangled [Fig.~\ref{fig:fig7}] localized spins of FM-analyzer. Beyond spintronics, STT-driven quantum dynamics of localized spins can be employed~\cite{Comas2019} to manipulate individual spin qubits and entangle them over very long distances.








\begin{acknowledgements}
We thank A. Suresh for technical help with TDNEGF+LLG-computed movies in the SM~\cite{sm} animating Fig.~\ref{fig:fig0}. M.~D.~P. and P.~P. were supported by ARO MURI Award No.~W911NF--14--0247. A.~E.~F. acknowledges support by the U.S. Department of Energy (DOE),  Office of Science, Basic Energy Sciences (BES) Grant No.~DE-SC0019275. P.~M. and B.~K.~N. were supported by the U.S. National Science Foundation (NSF) under Grant No. ECCS 1922689.
\end{acknowledgements}



	

\end{document}